\documentclass[letter]{aa} % for the letters 
\usepackage[]{hyperref}
\usepackage{graphicx}
\usepackage{txfonts}
\begin{document} 

   \title{The 3D structure of disc-instability protoplanets}

   %\subtitle{}

   \author{Adam Fenton
          \inst{1}
          \and
          Dimitris Stamatellos \inst{1}%\fnmsep\thanks{Just to show the usage of the elements in the author field}
          }

   \institute{Jeremiah Horrocks Institute for Mathematics, Physics, and Astronomy, University of Central Lancashire, Preston PR1 2HE, UK\\
              \email{dstamatellos@uclan.ac.uk}
             }

   \date{Received September 15, 2023; accepted December 16, 2023}

% \abstract{}{}{}{}{} 
% 5 {} token are mandatory
 
  \abstract
   {The model of disc fragmentation due to gravitational instabilities offers an alternate formation mechanism for gas giant planets, especially those on wide orbits.}
   {Our goal is to determine the 3D structure of disc-instability protoplanets and to examine how this relates to the thermal physics of the fragmentation process.}
   {We modelled the fragmentation of gravitationally unstable discs using the SPH code PHANTOM, and followed the evolution of the protoplanets formed  through the first and second-hydrostatic core phases (up to densities $10^{-3}\text{g\,cm}^{-3}$).}
   {We find that the 3D structure of disc-instability protoplanets is affected by the disc environment and the formation history of each protoplanet (e.g. interactions with spiral arms, mergers). The large majority of the protoplanets that form in the simulations are oblate spheroids rather than spherical, and they accrete faster from their poles. }
   {The 3D structure of disc-instability protoplanets is expected to affect their observed properties and should be taken into account when interpreting observations of protoplanets embedded in their parent discs.}


   \maketitle
%
%-------------------------------------------------------------------

\section{Introduction}

Disc fragmentation due to gravitational instabilities in relatively massive ($M_{\rm disc}\stackrel{>}{_\sim}0.1 M_\star$) protostellar discs \citep{Kuiper:1951a, Cameron:1978a,Boss:1997a, Rice:2003a, Stamatellos:2007c, Boley:2009a, Rice:2022b} provides an alternative mechanism to  core accretion \citep[e.g.][]{Goldreich:1973a, Drazkowska:2023i} for the formation of gas giant planets.

Gravitational instabilities develop in protostellar discs when  the Toomre criterion \citep{Toomre:1964a} is satisfied, 
\begin{equation}
  Q \equiv \frac{c_{s} (R) \kappa (R)}{\pi G \Sigma (R)} \stackrel{<}{_\sim}1,\,
\label{eqn:toomre}
\end{equation}
where $c_{\rm s}$ is the sound speed, $\kappa$ is the epicyclic frequency, and $\Sigma$ is the surface density of the disc, at an orbital radius $R$. Gravitational instability leads to disc fragmentation when the disc cools sufficiently quickly,  
$t_{\textup{cool}} < (0.5 - 2) t_{\textup{orb}}$ (i.e.  a few orbital
periods).   Magnetic fields may also play an important role in disc formation \citep{Wurster:2018a,Lebreuilly:2023v,Hennebelle:2020c} and subsequent disc fragmentation \citep{Commercon:2010a}. It is believed that magnetic fields tend to act towards suppressing disc fragmentation, although this may still happen under the appropriate conditions \citep{Commercon:2010a,Forgan:2017a,Deng:2021e}.

The fragments produced by gravitational instability have masses that are a  few times the mass of Jupiter (M$_{\rm J}$), but the final mass they acquire may be much higher 
\citep{Stamatellos:2009a, Kratter:2010a, Vorobyov:2013a, Kratter:2016a, Mercer:2017a, Fletcher:2019q}.  The disc instability theory naturally forms gas giant planets on wide orbits \citep{Stamatellos:2009a}, where both criteria for disc fragmentation are satisfied. However, interactions with passing stars may destroy an initial population of such planets \citep{Carter:2023g}, in line with direct imaging observations \citep[e.g.][] {Bowler:2018a, Vigan:2021w} that show that  massive gas giants on wide orbits are not very common (only a small percentage of stars host such planets, up to a maximum of  5 -10\% of stars, with a small dependence on the stellar host mass).

The evolution of disc-instability fragments to protoplanets goes though the phases of the first and second hydrostatic cores \citep{Larson:1969a, Masunaga:2000a, Stamatellos:2007b, Stamatellos:2009d}. This means that the initial stages of the evolution of disc-instability planets are similar to those of a star within a collapsing molecular cloud core, albeit at a much smaller scale: the initial core (i.e. the fragment formed by disc instability that will evolve to a planet) has a size of a few AU, has a mass of a few M$_{\rm J}$, and   is rather rapidly rotating \citep{Stamatellos:2009d,Mercer:2020a}. 

Observations of planets still embedded in their parent discs (usually referred to as protoplanets) have become possible in the last few years \citep[see review by][]{Currie:2023s}. The two protoplanets around the 5 Myr old star PDS~70 are the first unambiguous
discoveries \citep{Keppler:2018a,Haffert:2019d}. They orbit at distances of 20 and 34~AU from the central star.   PDS 70~b has an estimated mass of  $<12{\rm M_ J}$; the mass of PDS 70~c is uncertain.  These protoplanets show signatures of gas accretion, as evidenced by H$_\alpha$ emission \citep{Wagner:2018a,Haffert:2019d}, and are attended by
circumplanetary discs \citep{Stolker:2020p,Benisty:2021j}. Recently, a protoplanet has been discovered around Aurigae AB \citep{Currie:2022q}, a $1-3$~Myr old star. This protoplanet has an estimated mass of  $\sim 9~{\rm M_ J}$ and orbits at $\sim 93$~AU from its parent star.

 As more direct (and indirect) observations of protoplanets are likely in the near future, it is important to determine their properties when they form by different scenarios (core accretion and disc instability) so that we may identify the dominant gas giant planet formation mechanism. In this work we perform a set of hydrodynamic simulations of disc fragmentation to determine the 3D structure of disc-instability protoplanets. In Section 2 we discuss the disc initial conditions and the methods used for the simulations. In Section 3 we present their general results, and in Section 4 the density, temperature, and velocity profiles of the protoplanets that form in the simulations. In Section 5 we focus on the shape of disc-instability protoplanets, and in Section 6 we summarize the main results of our study.

\section{Methodology}
\label{sec:methods}

We model the thermodynamics of  gravitationally unstable discs with the smoothed particle hydrodynamics code PHANTOM \citep{Price:2018b}, using a barotropic equation of state \citep[e.g.][]{Bate:1998a}. We vary the density at which the equation of state switches from isothermal to adiabatic, the adiabatic index, and the initial disc temperature as this is set by stellar heating.

\subsection{Disc initial conditions}
We set a disc  with mass of $M_{\rm D}=0.6\,\rm{M}_{\odot}$ around a host star of 0.8$\,\rm{M}_{\odot}$.  The disc extends from $10-300$~AU and it is represented by $N_{\rm SPH}=4 \times 10^{6}$ particles. The disc mass is chosen so that many fragments can form due to disc fragmentation in each simulation and facilitate a statistical study of their properties. The minimum mass that can be resolved is $N_{\rm neigh} M_{\rm D}/N_{\rm SPH}\sim 7.5\times 10^{-4} {\rm M_J}$, which is much lower than the opacity limit for fragmentation \cite[a few ${\rm M_J}$; e.g.][]{Whitworth:2006a}. Therefore, disc fragmentation is property resolved.

The surface density of the disc is set to 
\begin{equation}
    \Sigma = \Sigma_{0}\left(\frac{R}{R_{\rm in}}\right)^{-3/2}(1-\sqrt{R_{\rm in}/R}),
    \label{equation:surface_density_power_law}
\end{equation}
where $R_{\rm in}=10$~AU is the inner disc radius, and $\Sigma_{0} = 1.53 \times 10^{3}\,\text{g\,cm}^{-2}$. The disc temperature profile is set to 
\begin{equation}
    T(R) = T_{{1\rm AU}}\left(\frac{R}{\rm{AU}}\right)^{-0.5}\, ,
    \label{eq:temperature-profile}
\end{equation}
where ${T}_{1\,\rm{AU}} = [150, 200]$~K. The above disc initial conditions ensure that the disc is initially Toomre unstable outside  $\sim 50$~AU.

\subsection{Disc thermodynamics}\label{section:disc_thermodynamics}

Hydrodynamic simulations often use a barotropic equation of state (i.e. $P\propto\rho^{\gamma}$) to reproduce the results of more computationally exhaustive radiative hydrodynamic simulations \citep{Masunaga:1998a,Masunaga:2000a,Whitehouse:2004a, Mercer:2018a}. To emulate the thermal effects during gravitational fragmentation in protostellar discs we use a hybrid four-piece barotropic equation of state  that is modified to include radiative feedback from the central protostar. More specifically  the temperature of an SPH particle $i$ is
\begin{equation}
T_i=\max\Bigl\{ T(R_i), T_B(\rho_i)\Bigr\} \, ,
\end{equation}
where $T(R_i)$ is set by the central star (see Eq.~\ref{eq:temperature-profile}) and $T_B(\rho_i)$ is provided from the barotropic equation,
\begin{equation}
  T_B (\rho)=
  \begin{cases}
                                T_0 & \text{, $\rho < \rho_{1}$} \\
                                T_0\left(\frac{\rho}{\rho_{1}}\right)^{(\gamma_{1}-1)}  &  \text{, $\rho_{1} \leq \rho <\rho_{2}$}\\
                                T_0\left(\frac{\rho_{2}}{\rho_{1}}\right)^{(\gamma_{1}-1)}\left(\frac{\rho}{\rho_{1}}\right)^{(\gamma_{2}-1)}  & \text{, $\rho_{2} \leq \rho <\rho_{3}$}\\
                                T_0\left(\frac{\rho_{2}}{\rho_{1}}\right)^{(\gamma_{1}-1)}\left(\frac{\rho_{3}}{\rho_{2}}\right)^{(\gamma_{2}-1)} \left(\frac{\rho}{\rho_{3}}\right)^{(\gamma_{3}-1)} & \text{, $\rho \geq \rho_{3}$},\\
                                
  \end{cases}
  \label{equation:barotropic_EOS}
\end{equation}
where $\gamma_{1}$, $\gamma_{2}$, and $\gamma_{3}$ are the adiabatic indices that control the stiffness of the equation of state in the three density regions (i.e. how fast the gas temperature rises due to compressional  heating during the collapse).
The first region 
($\rho<\rho_1$; typically $\rho_{1}\sim 10^{-13}\text{g\,cm}^{-3}$, $T\stackrel{<}{_\sim} 10$~K) 
corresponds to the phase of isothermal collapse where  the gas is optically thin and its radiation escapes freely. 
The second region 
($\rho_1<\rho<\rho_2$; typically $\rho_{2}\sim 3\times 10^{-12}\text{g\,cm}^{-3}$, $T\sim 10-100$~K, $\gamma=5/3$) corresponds to the phase where the gas is optically thick and starts heating up.  
The third region ($\rho_2<\rho<\rho_3$; typically $\rho_{3}\sim 6\times 10^{-9}\text{g\,cm}^{-3}$, $T\sim 100-2000$~K, $\gamma=7/5$) 
corresponds to the phase where the rotational degrees of molecular hydrogen have been excited. Finally,  the last region 
($\rho>\rho_3$; $T>2000$~K, $\gamma=1.1$) 
corresponds to the phase where molecular hydrogen starts to dissociate.

The first critical density, $\rho_1$, effectively determines when the fragment becomes optically thick, and is therefore  a measure of of the disc opacity and metallicity. The critical densities $\rho_2$ and $\rho_3$ are set (for each $\gamma_1$, $\gamma_2$) so as to correspond to temperatures of 100~K and 2,000~K, where the rotational degrees of molecular hydrogen are excited and the dissociation of molecular hydrogen commences, respectively.

 %______________________%
 \begin{figure}
     \centering    \includegraphics[width=0.9
     \columnwidth, keepaspectratio]{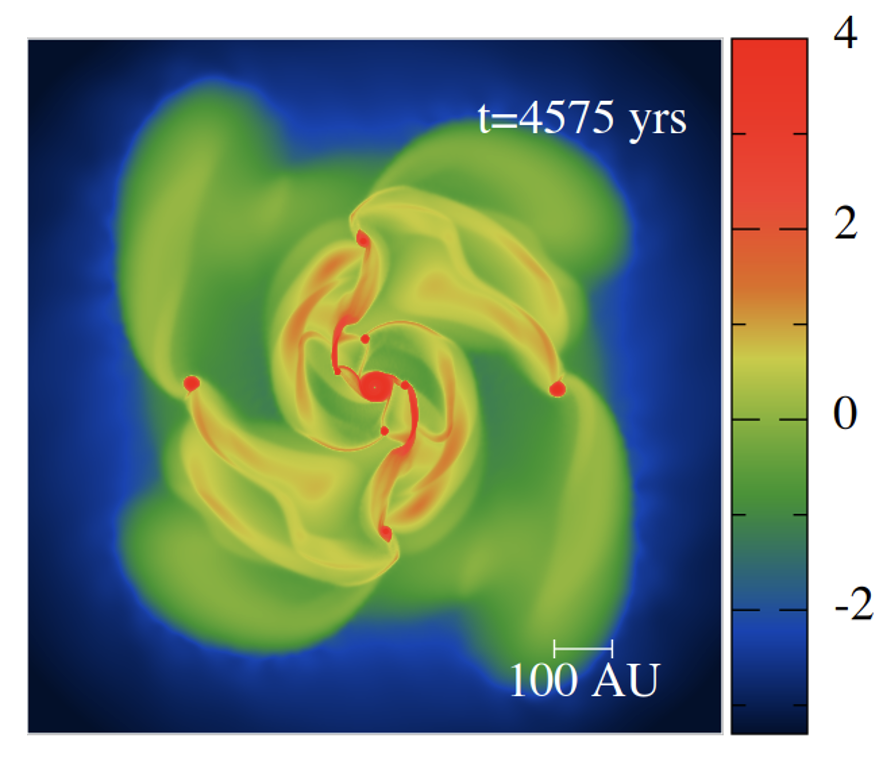}
     \caption{Surface density of the benchmark run disc (in $\text{g\,cm}^{-2}$). The disc becomes gravitationally unstable and fragments. Four of the fragments or protoplanets are followed until they reach density $10^{-3} \text{g\,cm}^{-3}$. }
          \label{fig:bm}
    \end{figure}  
  
\begin{figure}
\centering
\includegraphics[width=1\columnwidth, keepaspectratio]{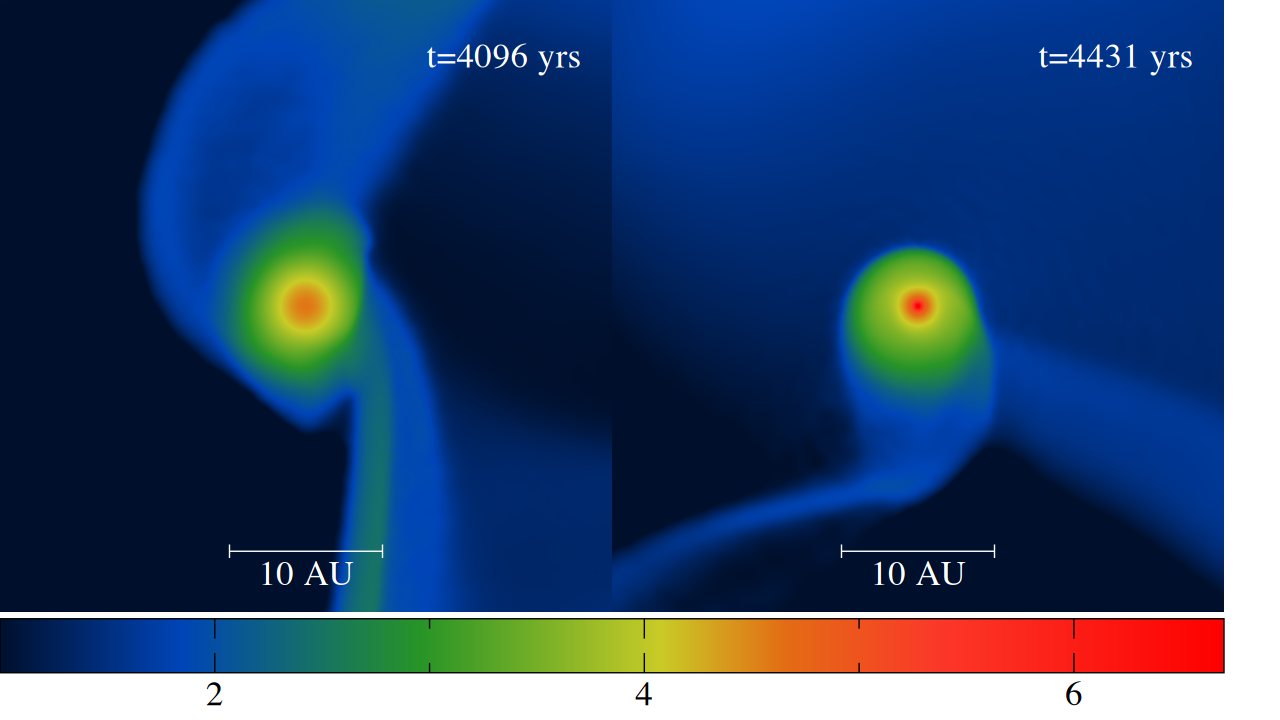}
\includegraphics[width=1\columnwidth, keepaspectratio]{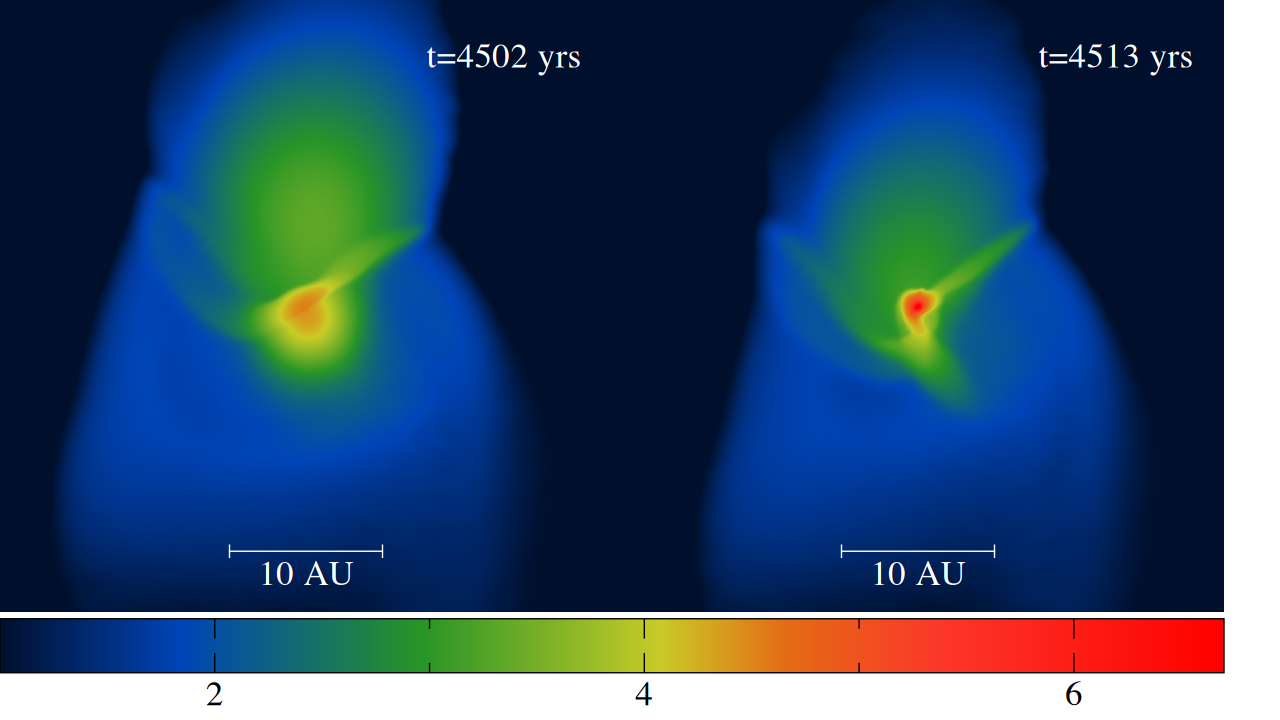}
     \caption{Surface density of two of the protoplanets (one per row; face-on view) that form in the benchmark run (in $\text{g\,cm}^{-2}$), plotted when their central density is  $\rho_{\rm c}$ = $10^{-9}$  (left column) and $\rho_{\rm c}$ = $10^{-5}\,\text{g\,cm}^{-3}$ (right column).}
     \label{fig:all_bm_clumps}
 \end{figure}
 %_____________________%
 
\section{Disc fragmentation and protoplanet formation}\label{section:the_parameter_space}

We performed high-resolution disc fragmentation simulations {with nine different sets of parameters. The parameter sets investigated are summarized in Table~\ref{tab:params}.} 
The disc initial conditions were chosen so that the discs quickly become gravitationally unstable, as evidenced by strong spiral arms, and fragment. These self-gravitating fragments are referred to as protoplanets.  We followed their  evolution to density  $10^{-3} \text{g\,cm}^{-3}$. Simulations with stiffer equations of state ($\gamma=1.66$) form fewer protoplanets, due increased compressional heating providing support against collapse. A total of 107 protoplanets form in all simulations.

A typical outcome of a simulation (the benchmark simulation) is shown in Fig.~\ref{fig:bm}. Four of the protoplanets that form in the simulation are followed until they reach a density of $10^{-3} \text{g\,cm}^{-3}$. Two of these protoplanets are shown in more detail in Fig.~\ref{fig:all_bm_clumps}, where they are plotted at two different times. {We also plot representative protoplanets, as seen face-on and edge-on, for each set of parameters (see Figs.~\ref{fig:all_clumps_xy}-\ref{fig:all_clumps2_xz}).
The non-axisymmetric and flattened morphology of these protoplanets is evident.}

%______________________%
 \begin{figure}
     \centering
\includegraphics[width=1.05\columnwidth, keepaspectratio]{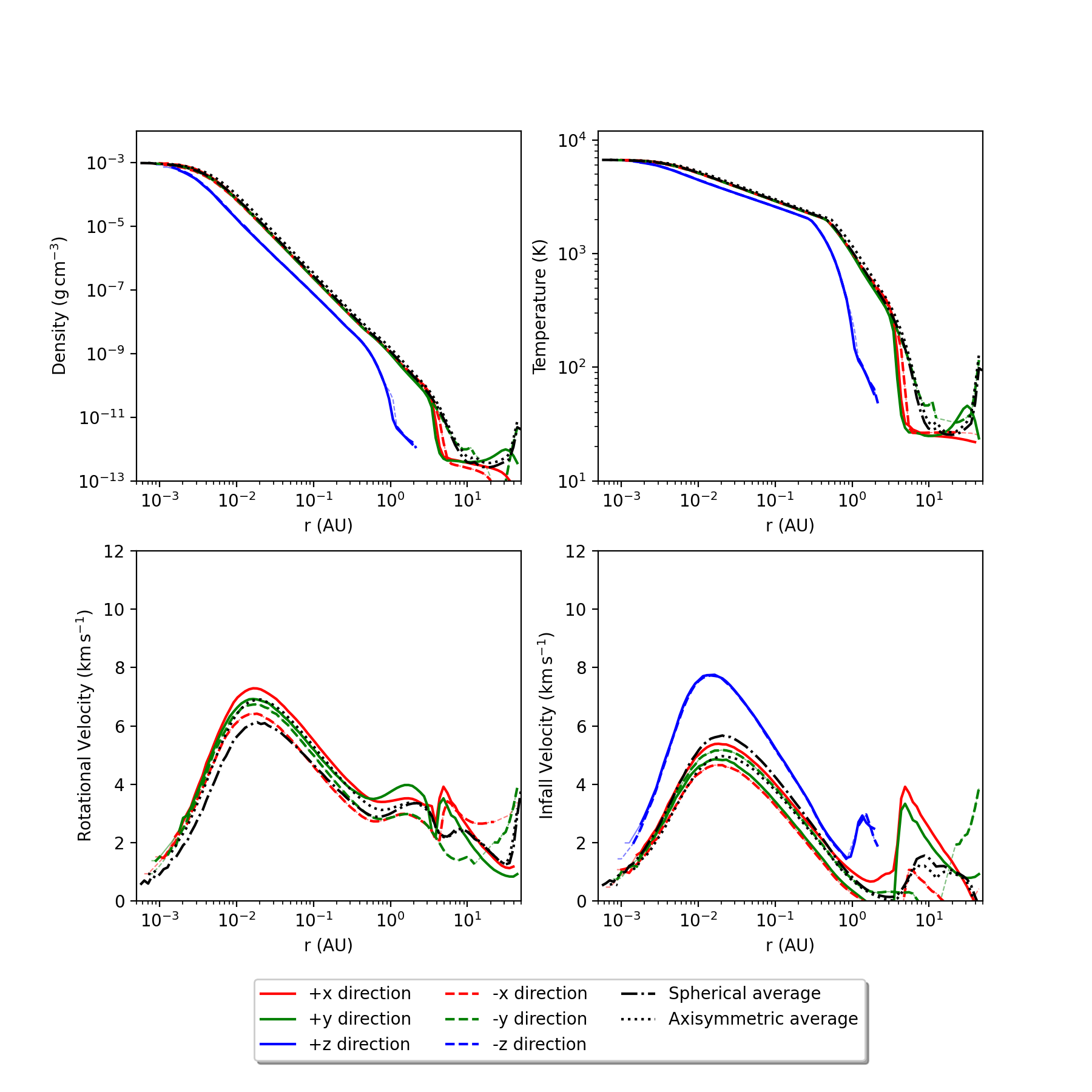}
     \caption{Density, temperature, rotational velocity, and infall velocity at different directions from the centre of one of the protoplanets that form in the benchmark run (see Fig.~\ref{fig:all_bm_clumps}, top) as marked on the graph. The axisymmetric averages are represented by the black dotted lines and the spherical averages are shown by the black dashed line.}
     \label{fig:BM_clump1_3D}
 \end{figure}
 %_____________________%
   %______________________%
 \begin{figure}
     \centering
     \includegraphics[width=1.01\columnwidth, keepaspectratio]{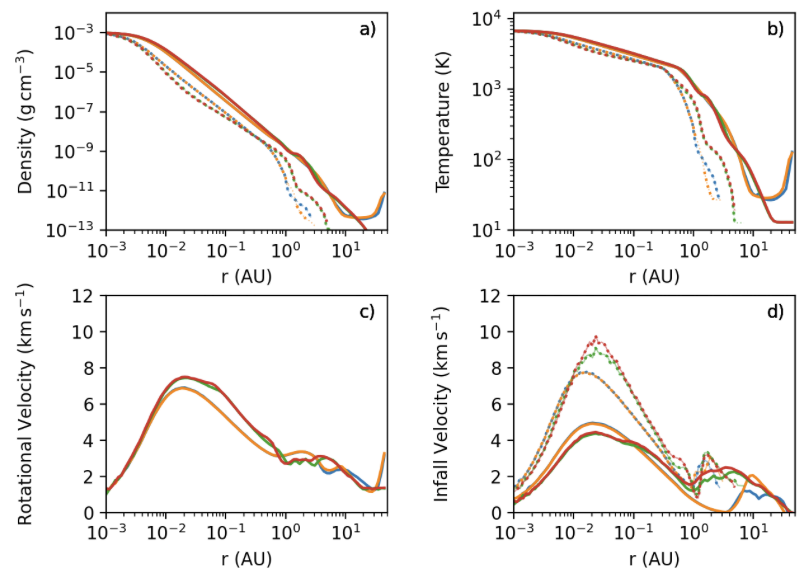}
     \caption{Radial profiles on the $x-y$ plane (solid lines) and vertical profiles ($\pm z$ direction; dotted lines) of the density, temperature, rotational, and infall velocity (panels a, b, c, and d, respectively) for the four protoplanets that form in the benchmark simulation.}
     \label{fig:BM_clump_profiles}
 \end{figure}
 %_____________________%

 The evolution of a fragment or protoplanet once it starts collapsing follows the same stages as the collapse of a solar-mass molecular cloud to a protostar \citep{Stamatellos:2009d}. The main difference is that the mass of the fragment itself is much lower, of the order of $10-100~{\rm M_ J}$. The collapse of the fragment or protoplanet is initially isothermal, with the temperature set by how far away the protoplanet is from the central star (typically $10-30$~K). Once the protoplanet becomes optically thick, the first hydrostatic core forms \citep{Larson:1969a, Stamatellos:2007b}, which grows in mass, and slowly contracts and heats up; an accretion shock forms around the first core as infalling gas decelerates. When the temperature rises to 2,000~K the molecular hydrogen dissociation initiates the second collapse and the second hydrostatic core forms \citep[see][]{Stamatellos:2009d, Mercer:2020a}. The mass of the first core is of the order of $10-20~{\rm M_J}$, whereas the mass of the second core is   a few ${\rm M_J}$. The final mass of the protoplanet will be decided through interactions with the disc \citep{Mercer:2020a}. Each protoplanet is represented by at least $6\times 10^5$ SPH particles, and therefore the thermodynamics of the collapse is properly resolved \citep{Stamatellos:2007b}.
%We follow the protoplanet evolution up to central densities $\rho = 10^{-3}\,\text{g\,cm}^{-3}$. At this point a sink particle is introduced with an accretion radius of 0.1$\,$AU to avoid small timesteps so that the evolution of the other protoplanets that form in the disc can be followed. 
The property of the protoplanets discussed later on refers to those when the density of $10^{-3}\,\text{g\,cm}^{-3}$ is reached at their centres.

\section{Three-dimensional structure of disc-instability protoplanets}

Previous studies \citep[e.g.][]{Mercer:2020a} have assumed that  disc-instability protoplanets,  are spherically symmetric. However, these  protoplanets   form  in a disc  with a nearly Keplerian rotational profile, such that they are rotating. Therefore, they are expected to be flattened, as   happens in rotating collapsing clouds leading to protostar formation \citep{Bate:1998a,Saigo:2006a,Saigo:2008a}.

%______________________%
 \begin{figure}
     \centering
     \includegraphics[width=0.9\columnwidth, keepaspectratio]{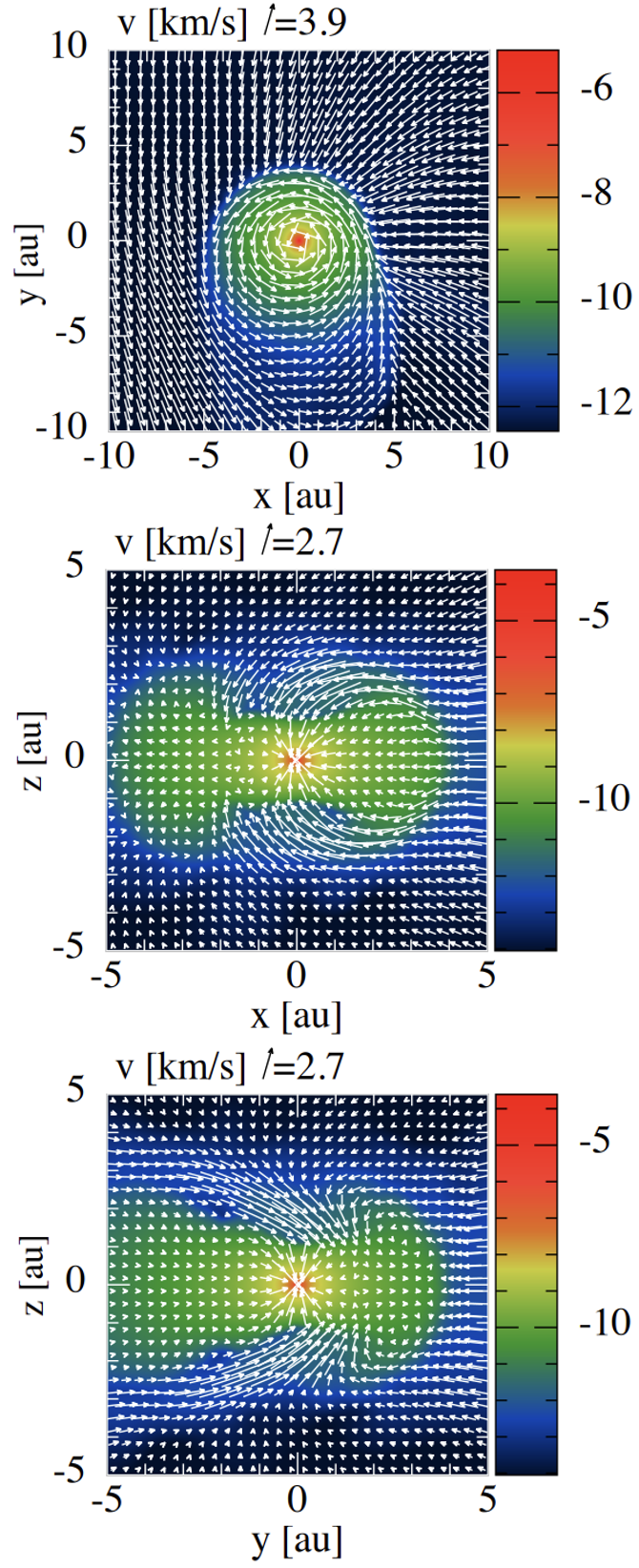}
     \caption{Velocity vectors of the gas flow onto a disc-instability protoplanet (see Fig.~\ref{fig:all_bm_clumps}, top) on the $x-y$ plane (top), $x-z$ plane (middle), and $y-z$ plane (bottom), overplotted on the corresponding densities (in units of $\text{g\,cm}^{-3}$). Gas infall velocities onto the protoplanet are asymmetric; higher velocities are seen towards the poles of the protoplanet.}
     \label{fig:vel}
 \end{figure}

 %______________________%
To investigate the 3D structure of the protoplanets formed in the simulations in more detail, we calculate the density, temperature, rotational velocity, and infall velocity along different directions from the centre ($\pm x$, $\pm y$, $\pm z$) of each protoplanet . We also calculate axisymmetric averages on the $x-y$ plane (which is assumed to be the plane of rotation) and spherical averages.
The results for a typical protoplanet are shown in Fig.~\ref{fig:BM_clump1_3D} (for the protoplanet formed in the benchmark simulation, see Fig.~\ref{fig:all_bm_clumps}a).
The density in the z-direction drops faster with radius than the density in the other two directions indicating that the protoplanet is flattened. The densities in the $\pm x$ and $\pm y$ directions are very similar apart from the edges of the protoplanet. This is also true for the temperature profile of the protoplanet. The rotational velocity of the protoplanet (which is not calculated in the z-direction, as this is the axis of rotation) shows differences along different directions as a result of the formation environment of the protoplanet that is being fed with gas from the disc. The infall velocity (see Fig.~\ref{fig:BM_clump1_3D}) is considerably higher along the poles of the protoplanet (i.e. in the $\pm z$ directions). The presence of the accretion shocks around the first and second core is evidenced by the maxima  and minima in the infall velocities (see Fig.~\ref{fig:BM_clump1_3D}), which correspond to the start of the shock when gas starts to decelerate before it falls onto the first and second core, respectively,  and stops.

In Fig.~\ref{fig:BM_clump_profiles} we compare the structure of the four protoplanets that form in the benchmark run. We plot axisymmetric averages (solid lines), and averages along the $\pm z$ direction (dotted lines) for the density, temperature, rotational velocity, and infall velocity. The four protoplanets show similar density and temperature profiles apart from the edges, where they interact with the protostellar disc. However, the rotational velocity and infall velocity of gas towards their centres show significant differences that are indicative of their different formation histories. In all cases the infall velocity along the poles (i.e. in the $\pm z$ directions) of the protoplanet is much higher (a factor of $\sim 2$) than that along the protoplanet equator (i.e. on the $x-y$ plane). This is demonstrated in Fig.~\ref{fig:vel} where the velocity vectors of the gas are plotted over the density at different planes for the protoplanet in Fig.~\ref{fig:all_bm_clumps}, top. We find that accretion of gas onto protoplanets happens from the polar directions, as is also expected for core-accretion planets \citep{Tanigawa:2012a}.

Our results for all protoplanets that form in the simulations show that (i) protoplanets are flattened and symmetric with respect to the $x-y$ plane (i.e. the protostellar disc midplane), and (ii) protoplanets are nearly axisymmetric, although there are differences near their edges due their formation environment { (e.g. interactions with spiral arms)} and formation history (i.e. when a protoplanet forms due to a collision between two fragments) { (see Figs.~\ref{fig:all_clumps_xy}-\ref{fig:all_clumps2_xz})}.
For simplicity, we   assume for the rest of the discussion that protoplanets are axisymmetric so that we can make comparisons between protoplanets formed in different simulations. We note that our study has neglected the effect of magnetic fields, which may influence disc formation and subsequent fragmentation \citep[e.g.][]{Commercon:2010a}, and therefore the 3D structure of protoplanets.

 \begin{figure}
     \centering
     \includegraphics[width=1.\columnwidth, keepaspectratio]{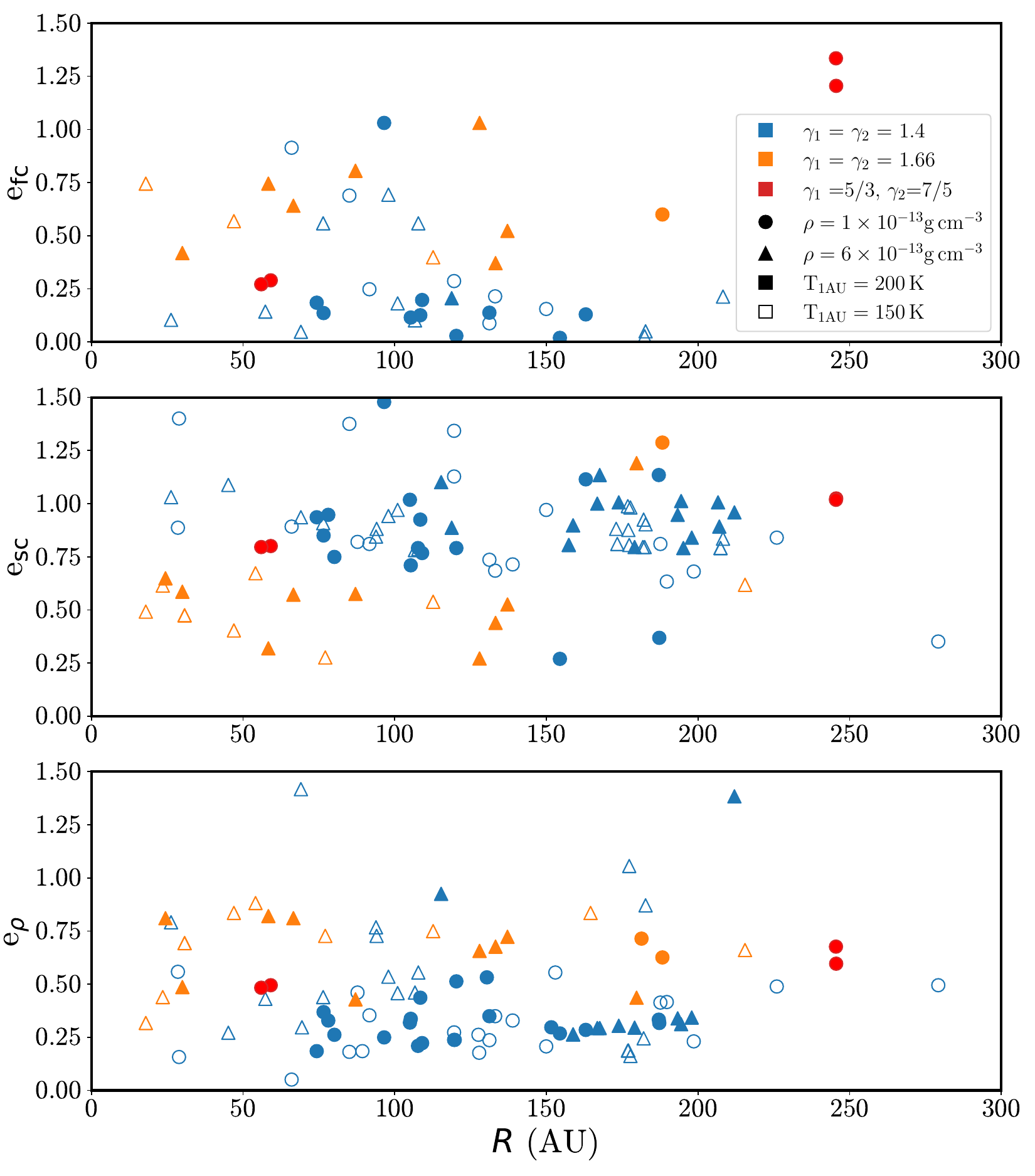}
     \caption{Aspect ratios of the first core $e_{\rm fc}$, the second core, $e_{\rm sc}$, and the fiducial core, $e_\rho$, (top to bottom) of all  protoplanets that form in the simulations. 
     The colour of each symbol corresponds to different adiabatic indices, the shape to different density threshold for the gas becoming optically thick (i.e to different disc opacities and metallicity), and the type of each symbol (filled or unfilled) to different stellar radiation fields (as marked on the graph legend).}
     \label{fig:aspect_ratios}
% \end{figure}
  %_____________________%
   %______________________%
% \begin{figure}
     \centering
      \includegraphics[width=1\columnwidth, keepaspectratio]{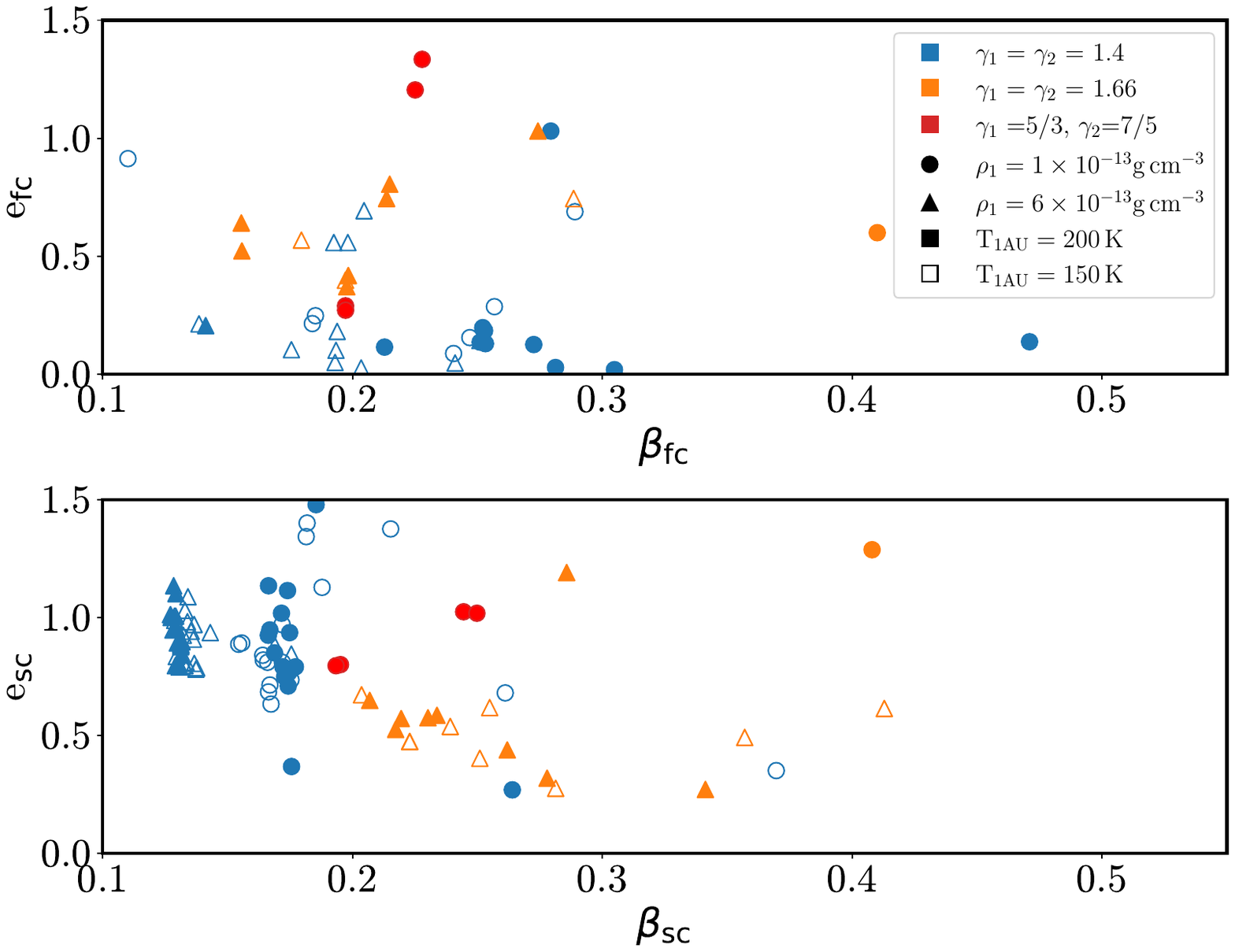}      
     \caption{Aspect ratios of the first and second core with respect to the ratios $\beta_{\rm fc}$, and $\beta_{\rm sc}$, of the rotational to the gravitational energy of the first and second core, respectively. Symbols as in Fig.~\ref{fig:aspect_ratios}. Second cores are generally more flattened when they rotate faster, but first cores do not show such dependence.}
     \label{fig:e-beta}
 \end{figure}
  %_____________________%

   %______________________%
 \begin{figure}
     \centering
      \includegraphics[width=1\columnwidth, keepaspectratio]{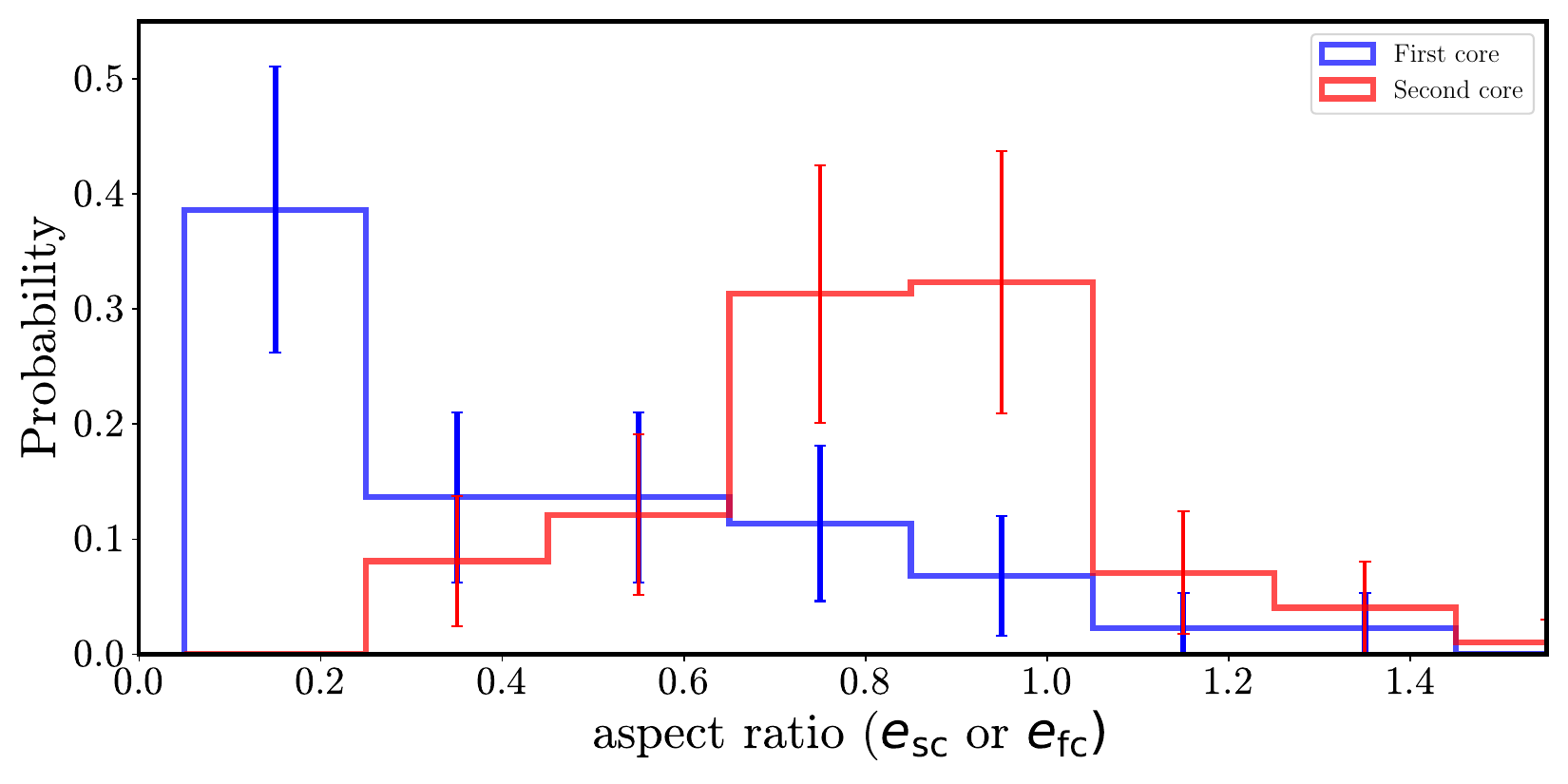} 
     \caption{Distribution of aspect ratios for the first (blue) and second cores (red). Most second cores are slightly flattened or nearly spherical  ($e_{\rm sc}\sim 0.7-1$), whereas first cores are highly flattened ($e_{\rm sc}\sim 0.1$) .}
     \label{fig:e-hist}
 \end{figure}
  %_____________________%

\section{The shape of disc-instability protoplanets}
\label{subsection:shape_of_fragments}

Disc-instability protoplanets are nearly axisymmetric (with respect to the rotation axis $z$) and symmetric with respect to the x-y plane; therefore, they can be  described as oblate spheroids.  To quantify their shape we use  three metrics. The first metric is the  aspect ratio of the first core, $e_{fc}$, as this is calculated at the inner boundary of the accretion shock around it (i.e. where the infall velocity is minimum or almost zero). This is the ratio of the inner first core radius on the $x-y$ plane over the corresponding radius in the $z$ direction (see Fig.~\ref{fig:BM_clump1_3D}). This ratio cannot be calculated accurately for all protoplanets as the outer region of the protoplanet is of low-density and therefore not well-resolved in some cases. The second  is the  aspect ratio of the second core, $e_{sc}$, as this is calculated at the outer boundary of the accretion shock around it (i.e. where the infall velocity is maximum). 
The third metric is the  aspect ratio, $e_\rho$, using a fiducial distance around the centre of the protoplanet where the density drops to $\rho_{\rm c} = 10^{-9}\text{g\,cm}^{-3}$ (which roughly corresponds to the radius of the first core), i.e. the ratio of the fiducial radius in the $z$ direction over the fiducial radius on the $x-y$ plane. This metric has the advantage that can be defined for all protoplanets.

 All three metrics paint the same picture regarding the protoplanets' morphology (see Fig~\ref{fig:aspect_ratios}).  Most protoplanets have aspect ratios $<1$, which means  they  are flattened, oblate spheroids rather than spherically symmetric. A few protoplanets have very high aspect ratios (above 1); these are possible outcomes of merger events.   
 
In  Fig.~\ref{fig:e-beta} we plot the aspect ratios of the the first and second core with respect to the corresponding ratios of the rotational to gravitational energy ($\beta_{\rm fc}$ and $\beta_{\rm sc}$, respectively). We see that second cores with higher $\beta_{\rm sc}$ values tend to be flat, but there is no such relation for the first cores. This suggests that the shape of the first cores is determined by interaction with the disc, whereas the shape of the second cores is due to their rotation. However, there are a few cases with high second core aspect ratio even with high rotational-to-gravitational energy ratios, suggesting a violent formation process (e.g. mergers or strong interactions with spiral arms).

 A comparison of the distributions of the aspect ratios of the first and second cores is shown in Fig.~\ref{fig:e-hist}. The two distributions are distinctly different, with second cores aspect ratios peaking around $e_{\rm sc}\sim 0.7-1$, whereas first cores are highly flattened, with aspect ratios peaking  around $e_{\rm fc}\sim 0.1$, which is similar to the disc scale height.
 
 A stiffer equation of state ($\gamma_1=1.66$ vs $\gamma_1=1.4$) generally  results in more spherical first cores ($\langle{e}_{\rm fc}\rangle = 0.62$ vs $\langle{e}_{\rm fc}\rangle=0.26$), but flatter second cores ($\langle e_{\rm sc}\rangle = 0.68 $ vs $\langle e_{\rm sc}\rangle = 0.96$; compare the different colours in Fig.~\ref{fig:aspect_ratios}). This occurs because a stiffer equation of state S also results in second cores with higher rotational-to-gravitational energy ratios ($\langle\beta_{\rm sc}\rangle=0.27$ vs $\langle\beta_{\rm sc}\rangle=0.17$; see Fig.~\ref{fig:e-beta}). For the first cores the $\beta$-ratios are similar $\langle\beta_{\rm fc}\rangle=0.23$.

A disc with higher metallicity and opacity
($\rho_{1}=6\times 10^{-13}\,\text{g\,cm}^{-3}$ vs $\rho_{1}=10^{-13}\,\text{g\,cm}^{-3}$)  results in slower rotating ($\langle\beta_{\rm fc}\rangle=0.19$ vs $\langle\beta_{\rm fc}\rangle=0.28$), but slightly more spherical first cores ($\langle e_{\rm fc}\rangle =0.41$ vs $\langle e_{\rm fc}\rangle = 0.38$), and in slower rotating ($\langle\beta_{\rm sc}\rangle=0.17$ vs $\langle\beta_{\rm sc}\rangle=0.21,$ but slightly flatter second cores ($\langle e_{\rm sc}\rangle = 0.84$ vs $\langle e_{\rm sc}\rangle = 0.99$). However, these differences are rather small (compare triangles with circles in Figs.~\ref{fig:aspect_ratios}-\ref{fig:e-beta}).
A disc with greater heating from the central star 
($T_{1{\rm AU}}=200$~K vs $T_{1{\rm AU}}=150$~K) has no effect on the rotational-to-gravitational energy ratios, but the first and second cores are slightly more spherical ($\langle e_{\rm fc}\rangle = 0.46$ vs $\langle e_{\rm fc}\rangle = 0.33 $, $\langle e_{\rm sc}\rangle = 0.92 $ vs $\langle e_{\rm sc}\rangle =0.87$, respectively)  (compare filled with unfilled symbols in Figs.~\ref{fig:aspect_ratios}-\ref{fig:e-beta})).
 Finally, there seems to be no correlation between the shapes of the first and second cores and the position where they form within the disc.

\section{Conclusions}

Disc-instability protoplanets are not spherically symmetric, but close to being oblate spheroids. Their outer regions show more complex, asymmetric structure due to interactions with the protostellar disc and their formation history. Gas accretion happens faster from the protoplanet poles than from the protoplanet equator. We expect that this may lead to a strong modification  of the observed properties of  protoplanets \citep[e.g. their spectrum, H$_\alpha$ emission; see][]{Zhu:2015a,Marleau:2022v,Marleau:2023l} with the viewing angle that needs to be taken into account when interpreting observations, like those of PDS~70~b,c \citep{Keppler:2018a,Haffert:2019d} and AB~Aurigae~b \citep{Currie:2022q}.

\begin{acknowledgements}
{ We thank the anonymous referee for suggestions that helped improving the paper. } The simulations were performed using the UCLan High Performance Computing (HPC) facility, the DiRAC Memory Intensive service at Durham University, managed by the Institute for Computational Cosmology on behalf of the STFC DiRAC HPC Facility (www.dirac.ac.uk), and the DiRAC Data Intensive service at the University of Leicester, managed by the University of Leicester Research Computing Service.  The DiRAC service  was funded by BEIS, UKRI and STFC capital funding and STFC operations grants. DiRAC is part of the UKRI Digital Research Infrastructure. Surface density plots were produced using SPLASH \citep{Price:2007b}.  We acknowledge support from STFC grant ST/X508329/1.

\end{acknowledgements}

%-------------------------------------------------------------------
\bibliography{bibliography.bib} 

\appendix

\section{Simulation parameters and gallery of protoplanet surface density plots}

The parameter sets investigated are listed in Table~\ref{tab:params}. For an explanation of the different parameters see Section~\ref{sec:methods}.

\begin{figure}
\centering
\includegraphics[width=1\columnwidth, keepaspectratio]{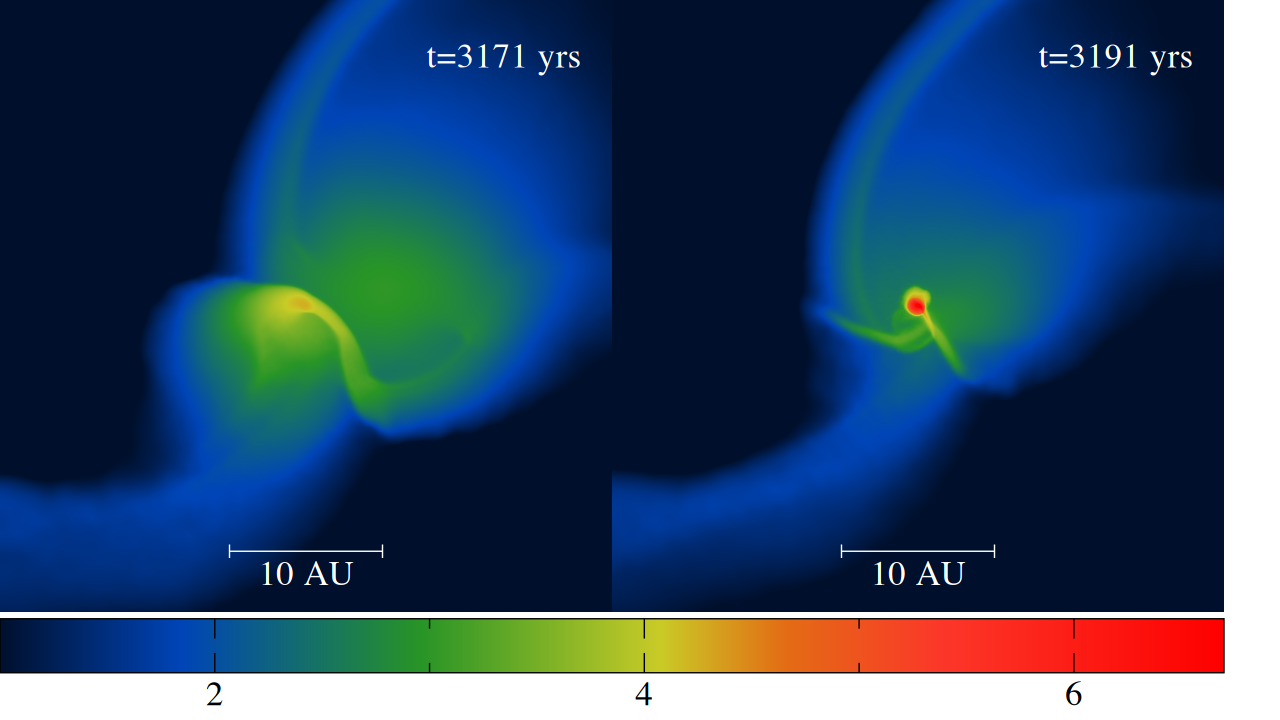}
\includegraphics[width=1\columnwidth, keepaspectratio]{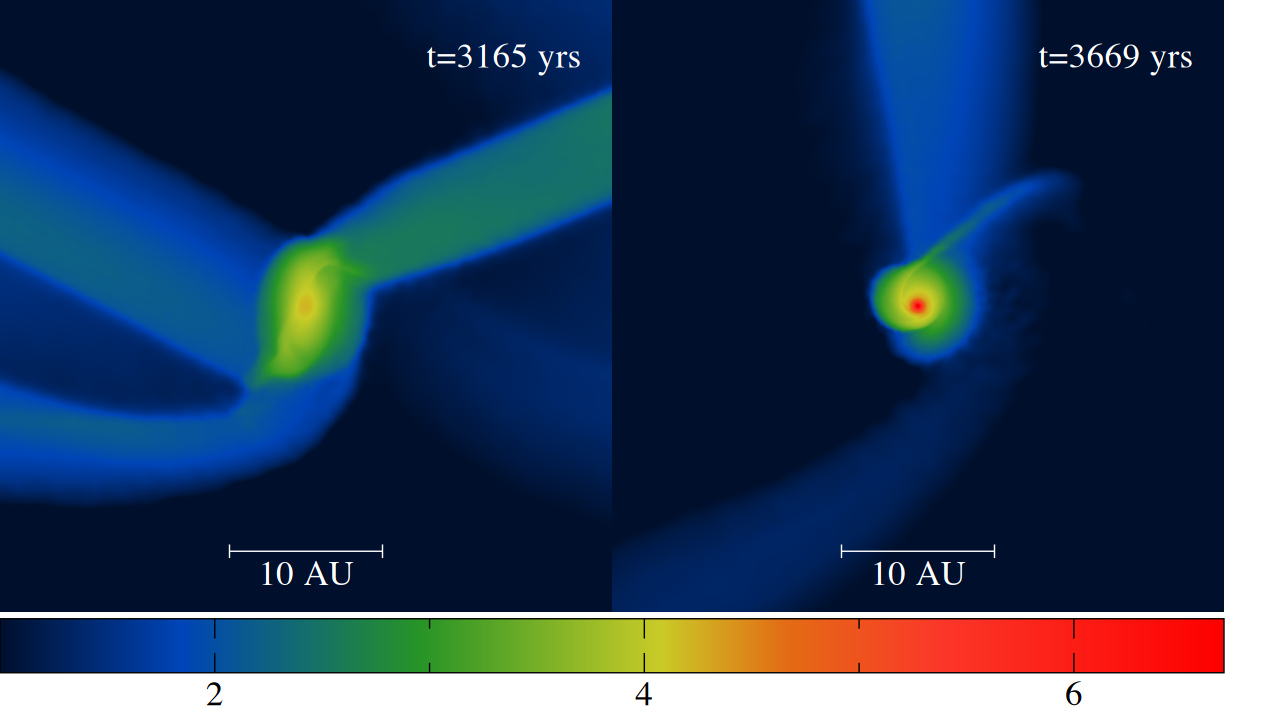}
\includegraphics[width=1\columnwidth, keepaspectratio]{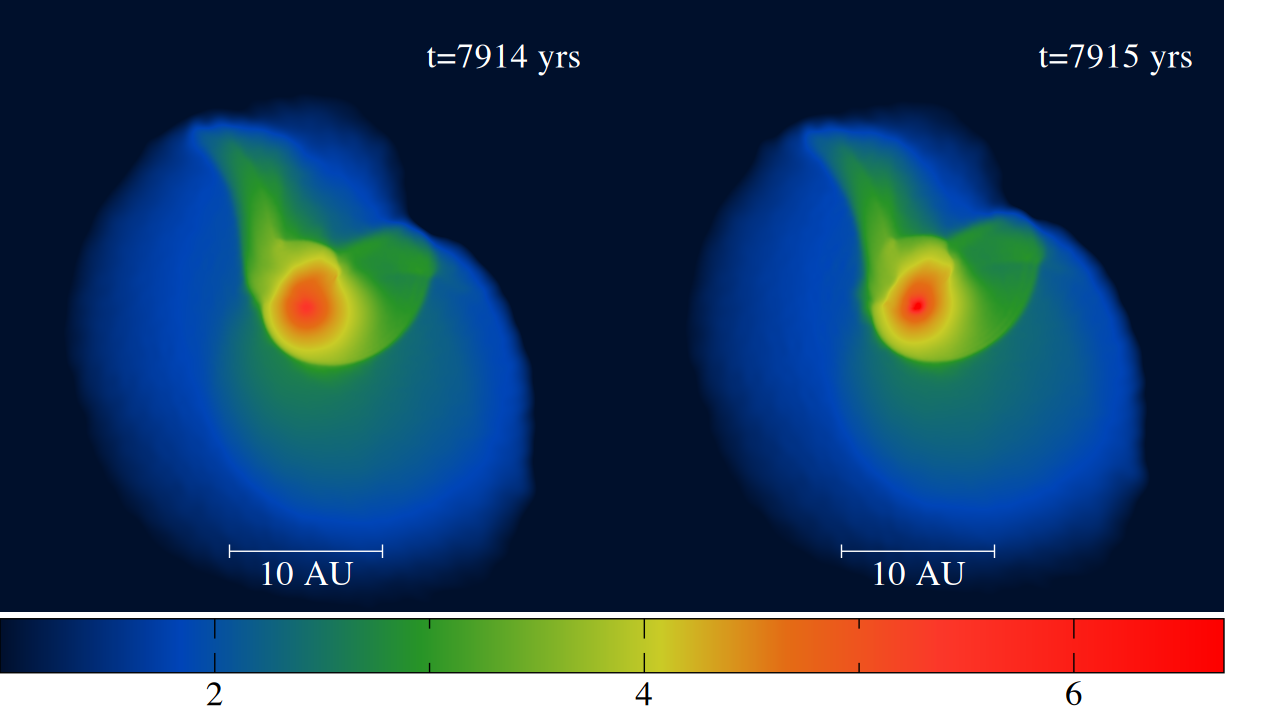}
\includegraphics[width=1\columnwidth, keepaspectratio]{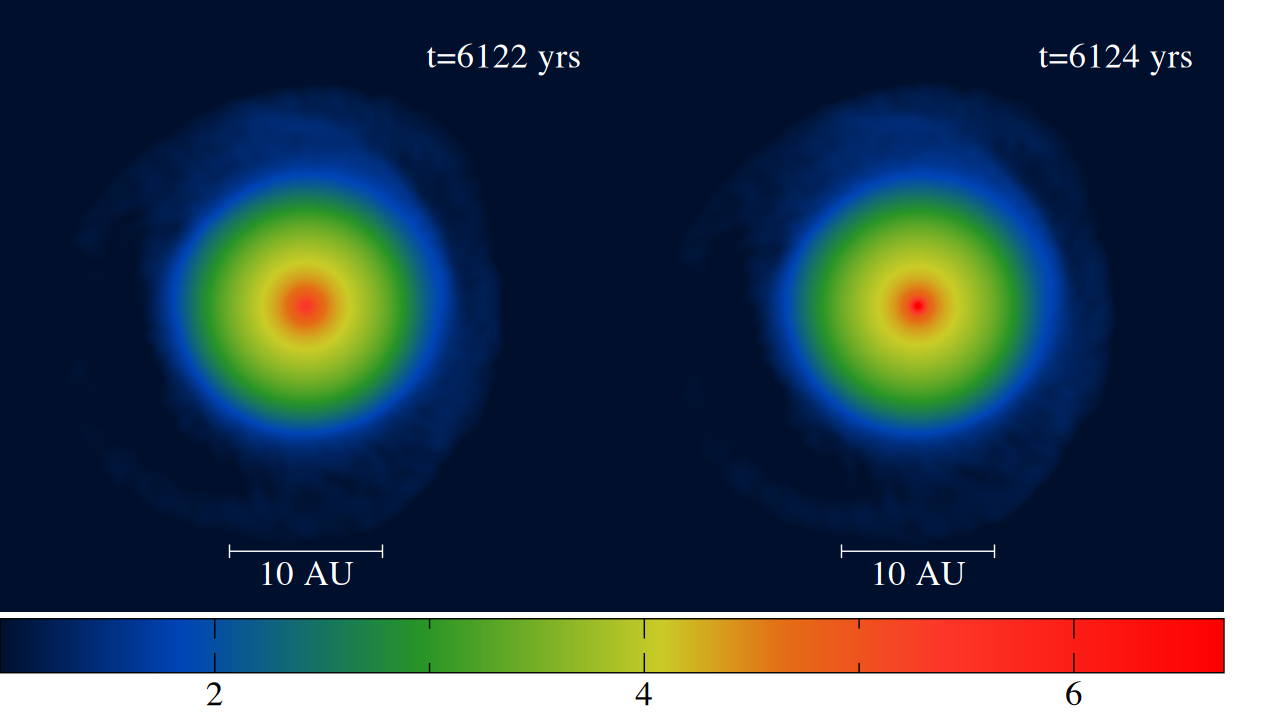}
    \caption{Surface density of representative protoplanets (face-on view) that form in Runs $1-4$ (from top to bottom), in $\text{g\,cm}^{-2}$, plotted when their central density is  $\rho_{\rm c}$ = $10^{-9}$  (left column) and $\rho_{\rm c}$ = $10^{-5}\,\text{g\,cm}^{-3}$ (right column).}
     \label{fig:all_clumps_xy}
 \end{figure}

\begin{figure}
\centering
\includegraphics[width=1\columnwidth, keepaspectratio]{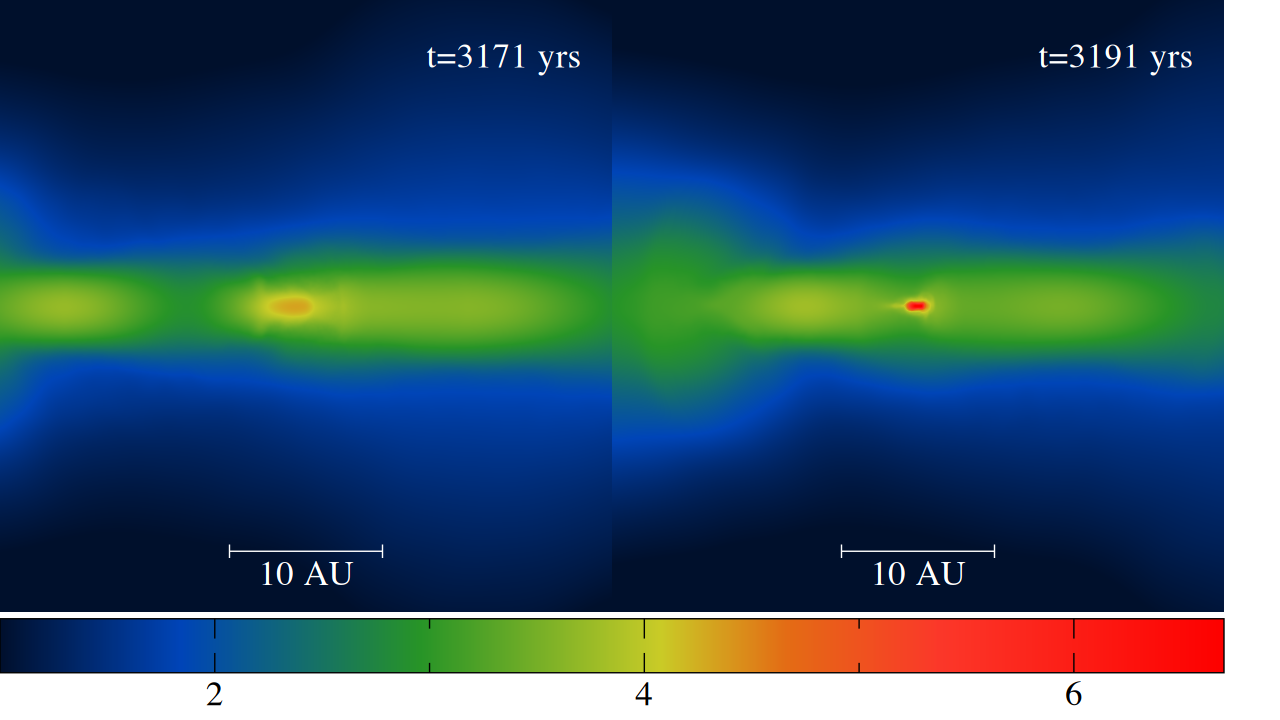}
\includegraphics[width=1\columnwidth, keepaspectratio]{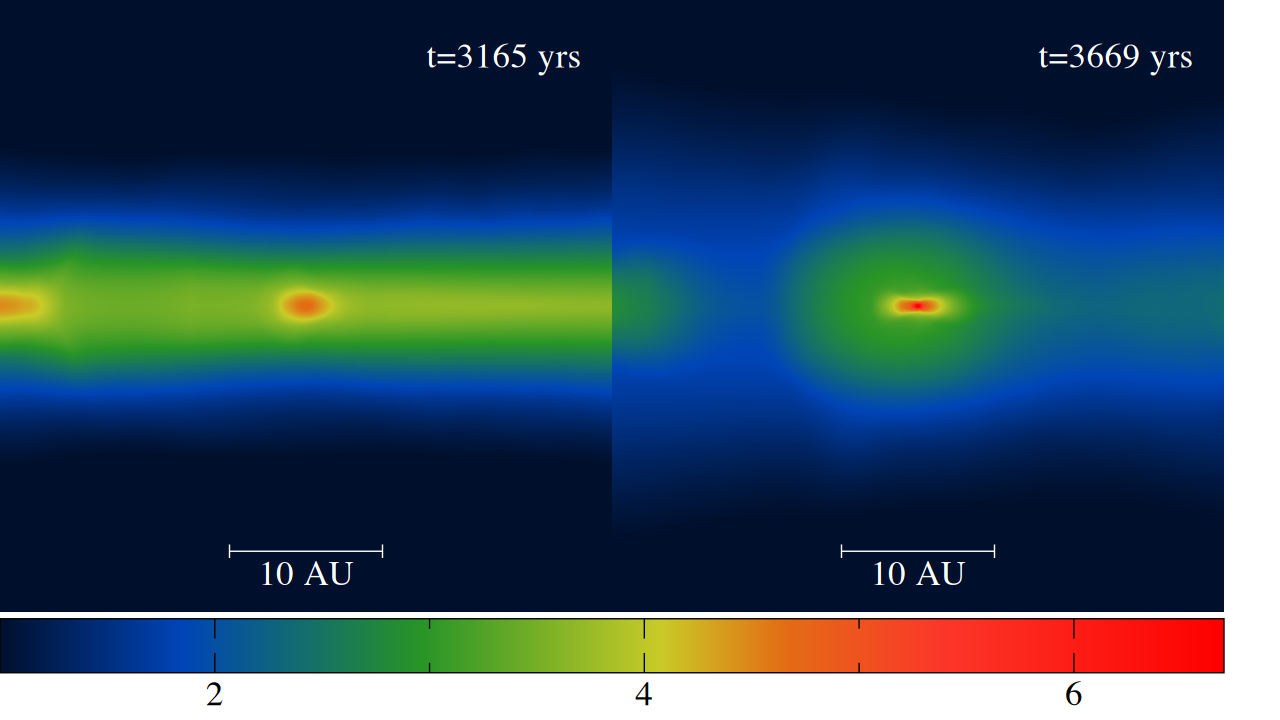}
\includegraphics[width=1\columnwidth, keepaspectratio]{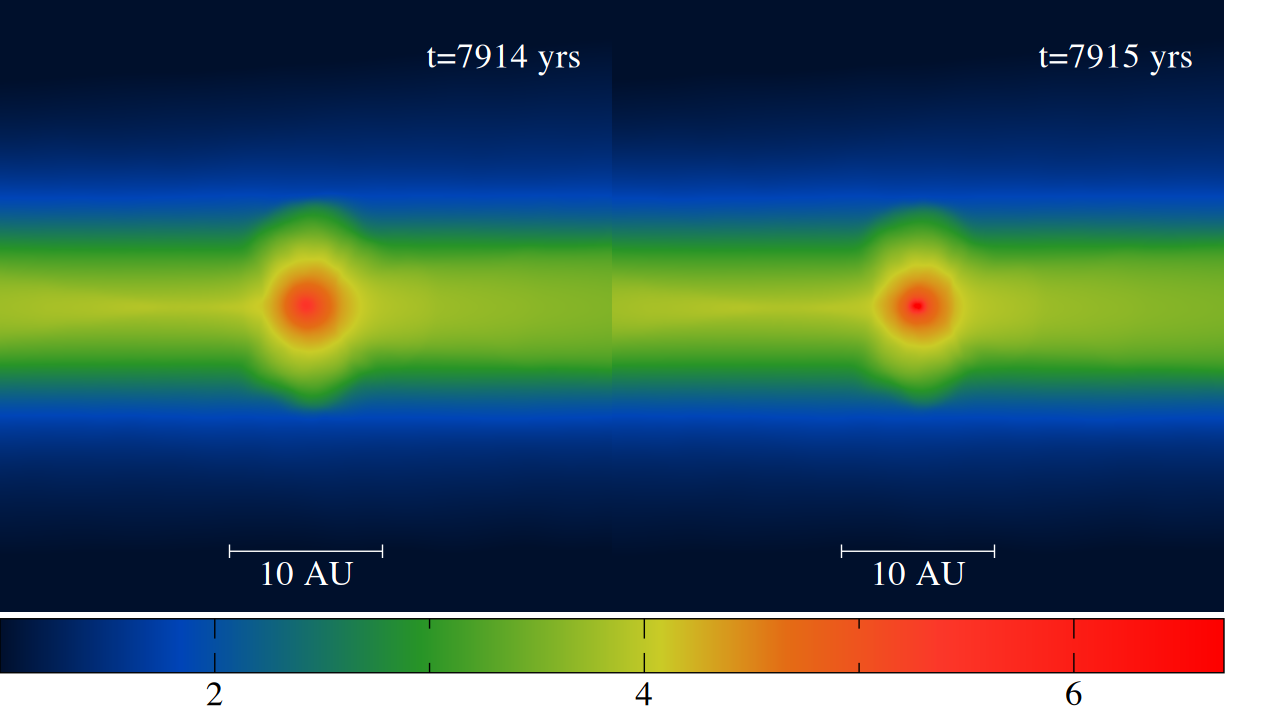}
\includegraphics[width=1\columnwidth, keepaspectratio]{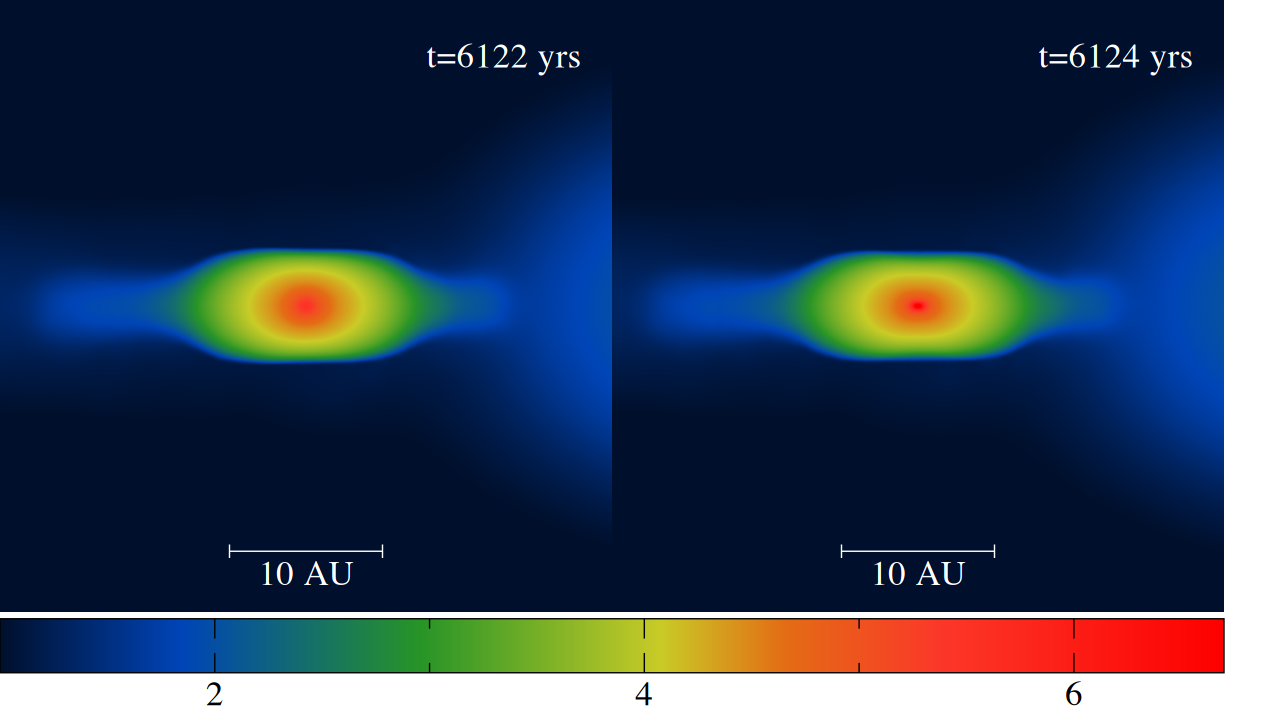}
      \caption{Surface density of representative protoplanets (edge-on view) that form in Runs $1-4$ (from top to bottom), in $\text{g\,cm}^{-2}$, plotted when their central density is  $\rho_{\rm c}$ = $10^{-9}$  (left column) and $\rho_{\rm c}$ = $10^{-5}\,\text{g\,cm}^{-3}$ (right column).}
     \label{fig:all_clumps_xz}
 \end{figure}

\begin{figure}
\centering
\includegraphics[width=1\columnwidth, keepaspectratio]{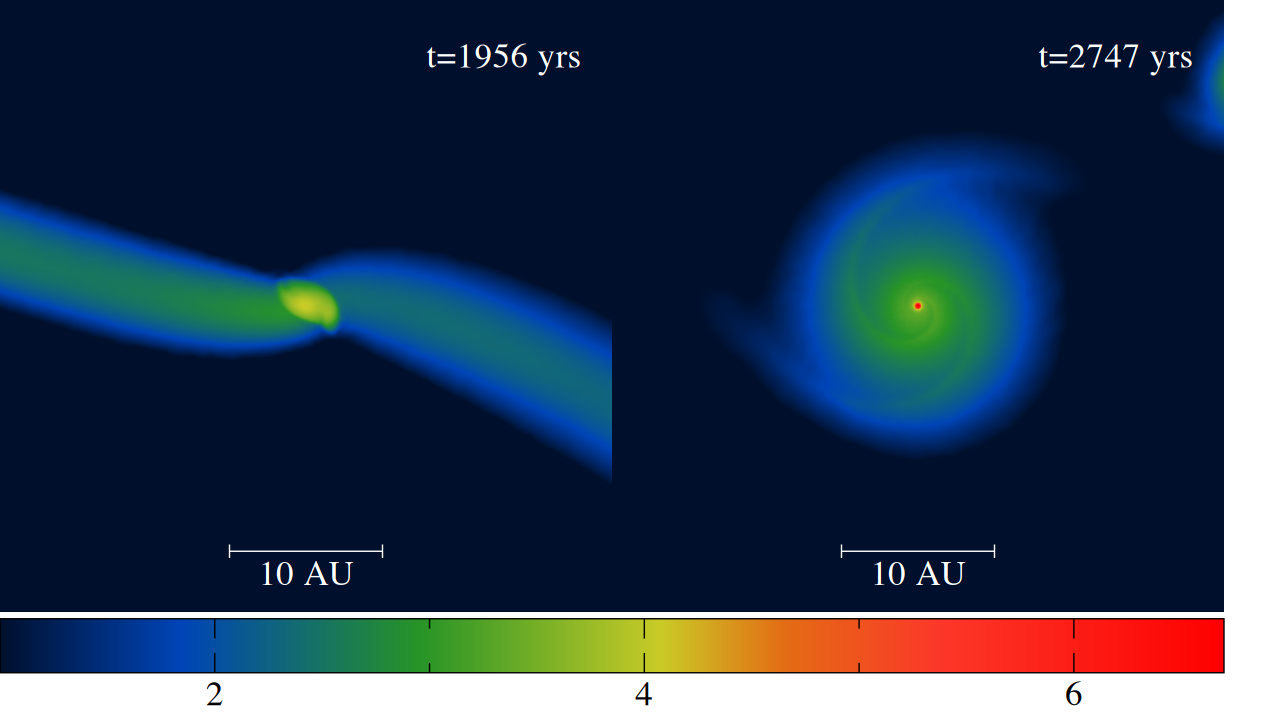}
\includegraphics[width=1\columnwidth, keepaspectratio]{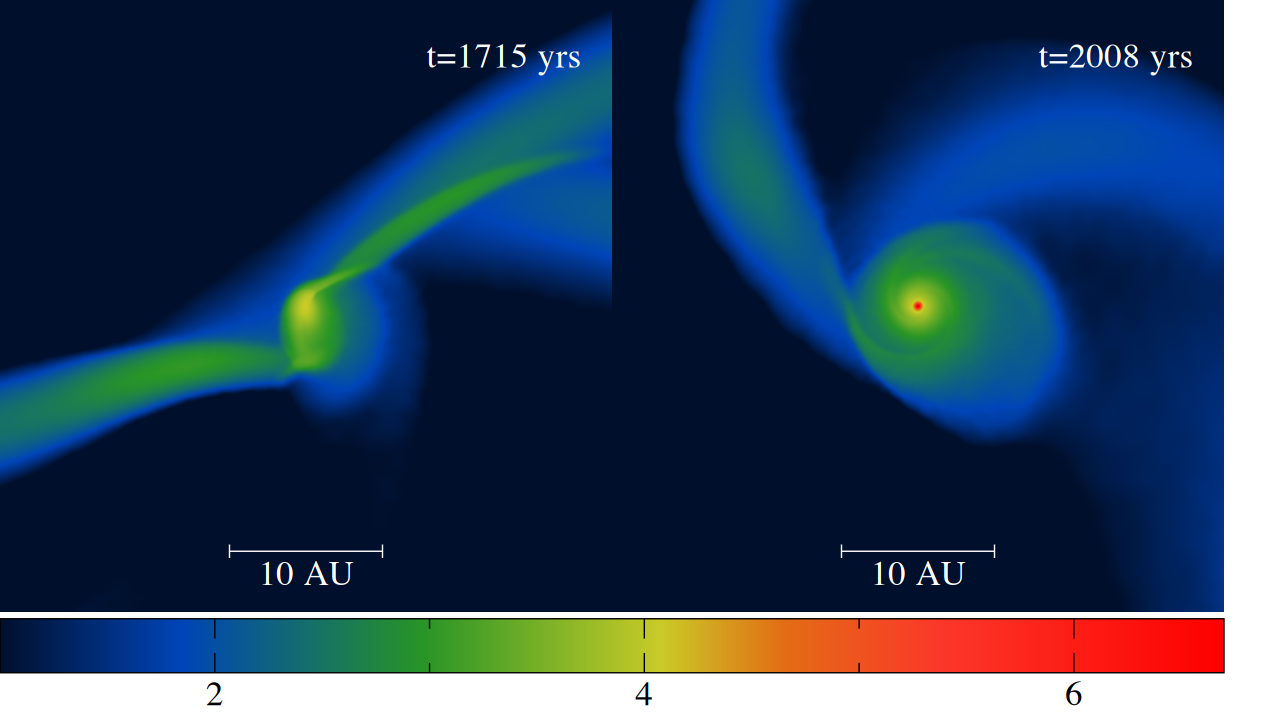}
\includegraphics[width=1\columnwidth, keepaspectratio]{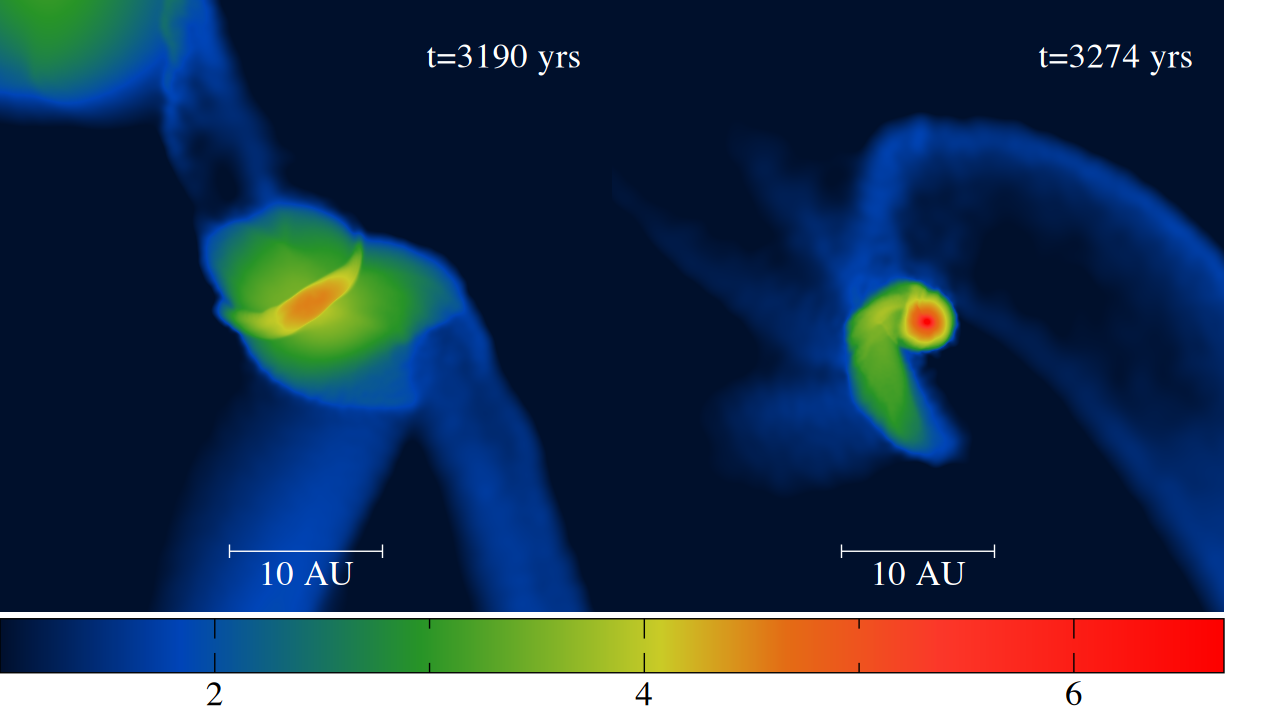}
\includegraphics[width=1\columnwidth, keepaspectratio]{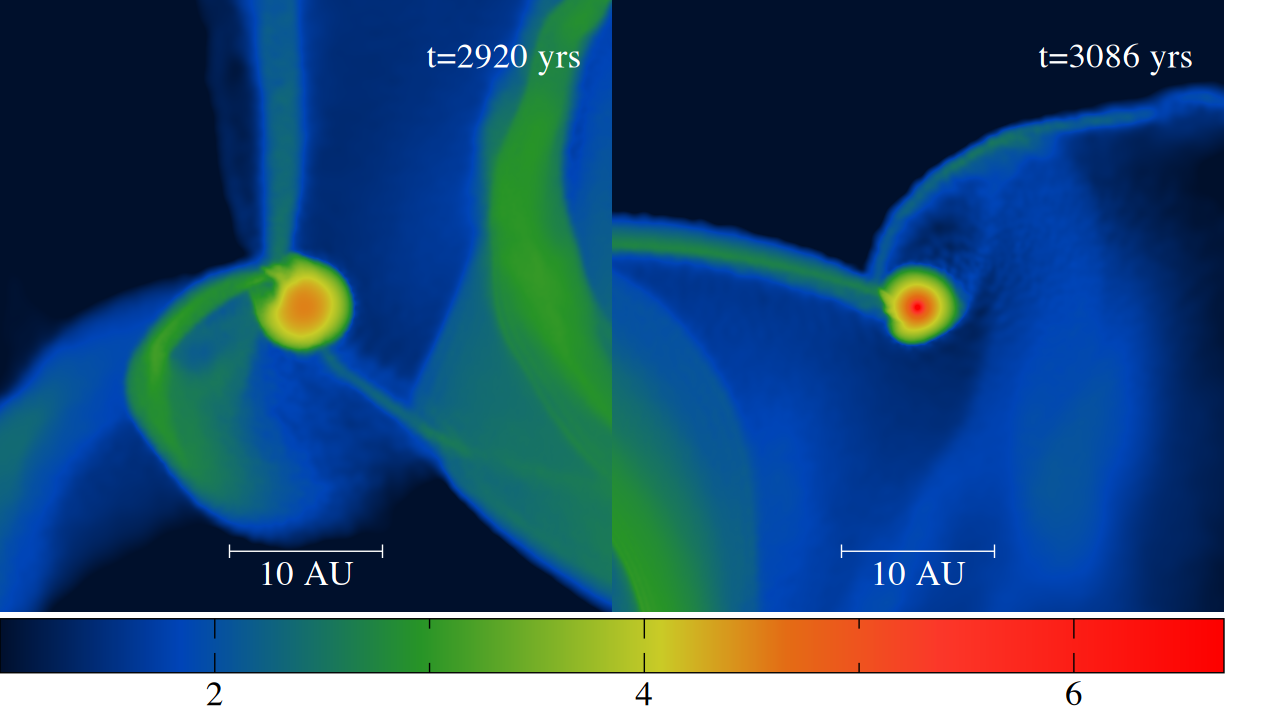}
     \caption{Same as in Fig.~\ref{fig:all_clumps_xy}, but for protoplanets in Runs $5-8$.}
     \label{fig:all_clumps2_xy}
 \end{figure}

\begin{figure}
\centering
\includegraphics[width=1\columnwidth, keepaspectratio]{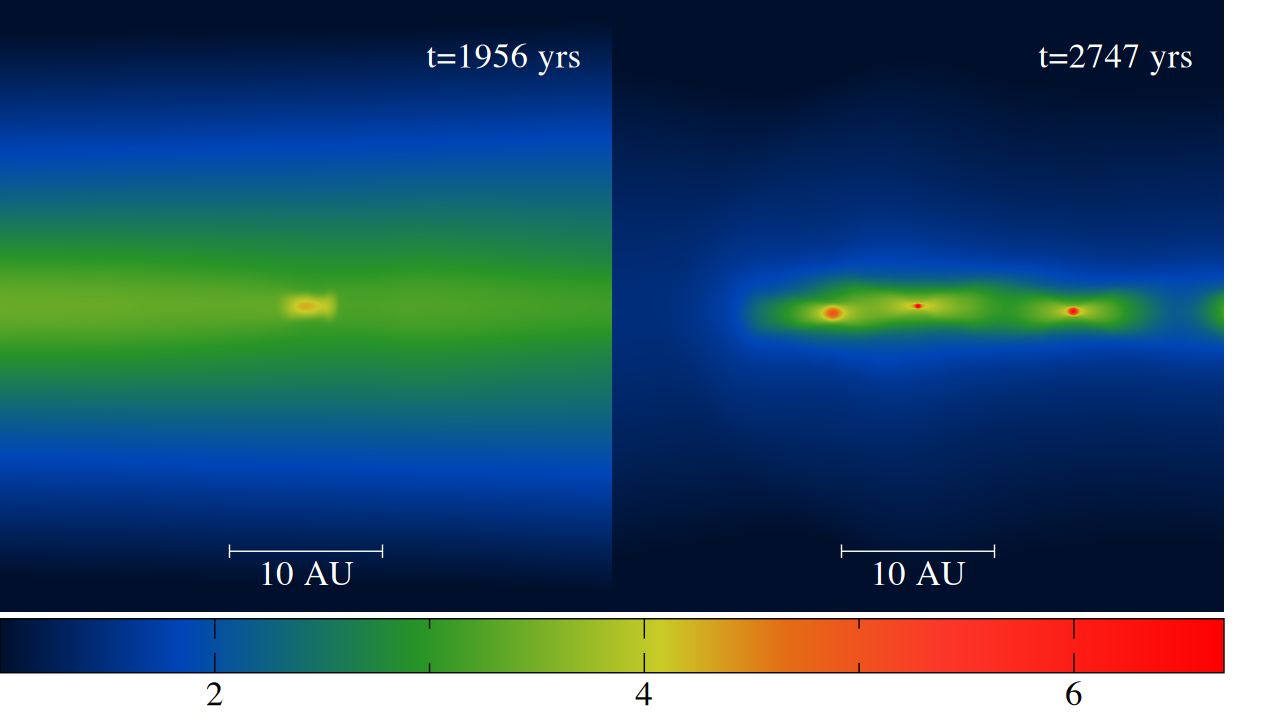}
\includegraphics[width=1\columnwidth, keepaspectratio]{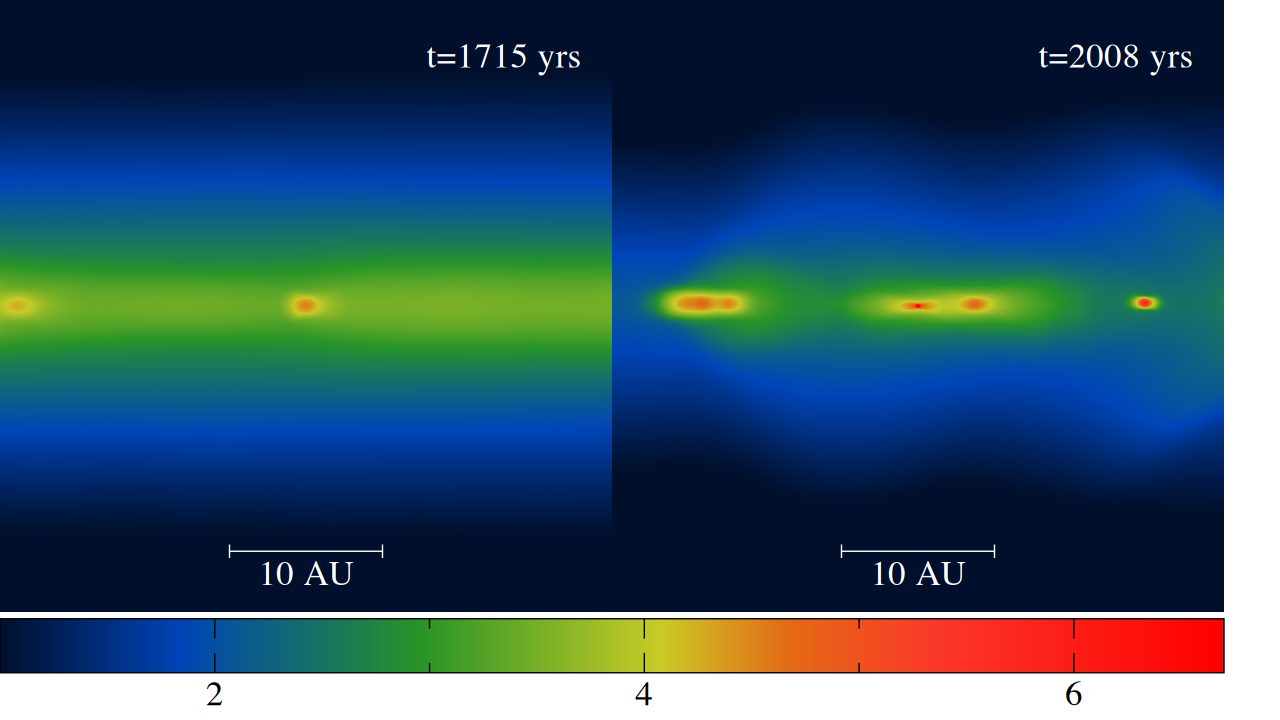}
\includegraphics[width=1\columnwidth, keepaspectratio]{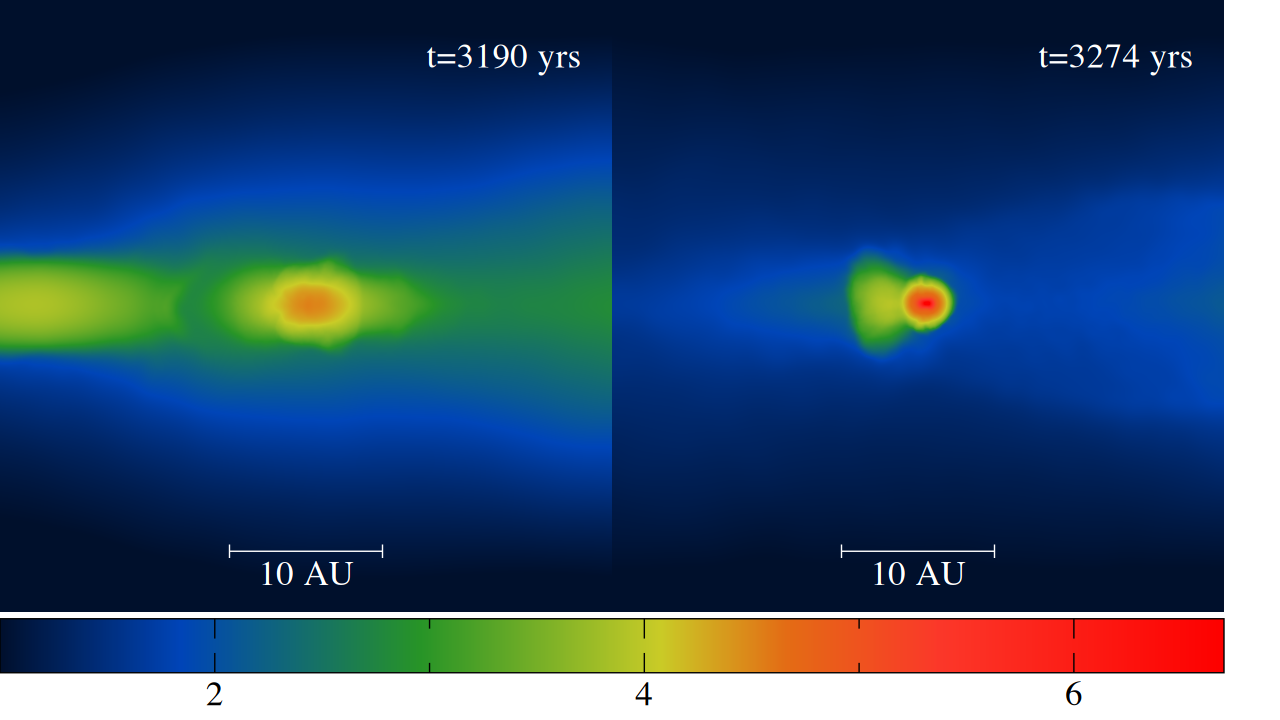}
\includegraphics[width=1\columnwidth, keepaspectratio]{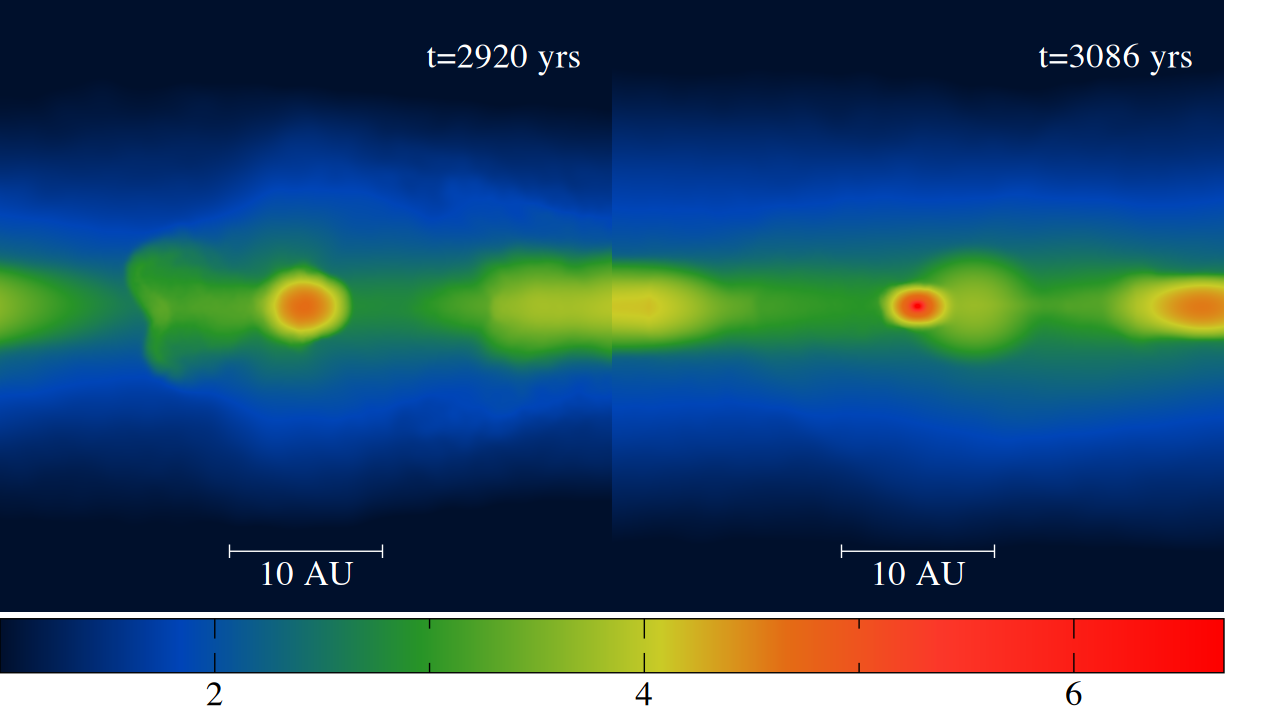}
     \caption{Same as in Fig.~\ref{fig:all_clumps_xz}, but for protoplanets in Runs $5-8$. Due to the projection, more than one protoplanet appears in some of these plots.}
     \label{fig:all_clumps2_xz}
 \end{figure}

Representative protoplanets for each of the eight sets of equation of state parameters are plotted in Figs.~\ref{fig:all_clumps_xy}-\ref{fig:all_clumps2_xz}. The different morphologies of the protoplanets for different equation of state parameters are clearly seen in these plots.

\begin{table*}
\centering
\caption{Equation of state parameters used for the disc fragmentation simulations (see Section~\ref{sec:methods}).}

\begin{tabular}[ht]{cccccccc}
\hline
ID&$\rho_{1}\,(\text{g\,cm}^{-3})$&$\rho_{2}\,(\text{g\,cm}^{-3})$&$\rho_{3}\,(\text{g\,cm}^{-3})$&$\gamma_{1}$&$\gamma_{2}$&$\gamma_{3}$&$\rm{T}_{1\,\rm{AU}} (\rm K)$\\
\hline
Benchmark& $1\times 10^{-13}$&$3.27\times10^{-12}$&$5.86\times 10^{-9}$&1.66&1.4&1.1&200.0\\
Run 1&  $1\times  10^{-13}$&   $3.16 \times 10^{-11}$  &$5.66\times 10^{-8}$   &1.4&1.4&1.1&200.0\\
Run 2&  $1\times  10^{-13}$&   $3.16 \times 10^{-11}$  &$5.66\times 10^{-8}$   &1.4&1.4&1.1&150.0\\
Run 3&  $1\times  10^{-13}$&   $3.27 \times 10^{-12}$  &$3.06\times 10^{-10}$  &1.66&1.66&1.1&200.0\\
Run 4&  $1\times  10^{-13}$&   $3.27 \times 10^{-12}$  &$3.06\times 10^{-10}$  &1.66&1.66&1.1&150.0\\
%Run 5&  $1\times  10^{-13}$&   $1.78 \times 10^{-12}$  &$7.52\times 10^{-11}$  &1.8&1.8&1.1&200.0\\
%Run 6&  $1\times  10^{-13}$&   $1.78 \times 10^{-12}$  &$7.52\times 10^{-11}$  &1.8&1.8&1.1&150.0\\
%Run 7&  $1\times  10^{-13}$&   $1    \times 10^{-8}$  &$3.20\times 10^{-2}$              &1.2&1.2&1.1&200.0\\
%Run 8&  $1\times  10^{-13}$&   $1    \times 10^{-8}$  &$3.20\times 10^{-2}$    &1.2&1.2&1.1&150.0\\
Run 5&  $6\times  10^{-13}$&   $1.9  \times 10^{-10}$  &$3.39\times 10^{-7}$   &1.4&1.4&1.1&200.0\\
Run 6& $6\times  10^{-13}$&   $1.9  \times 10^{-10}$  &$3.39\times 10^{-7}$   &1.4&1.4&1.1&150.0\\
Run 7& $6\times  10^{-13}$&   $1.96 \times 10^{-11}$  &$1.84\times 10^{-9}$   &1.66&1.66&1.1&200.0\\
Run 8& $6\times  10^{-13}$&   $1.96 \times 10^{-11}$  &$1.84\times 10^{-9}$   &1.66&1.66&1.1&150.0\\
%Run 13& $6\times  10^{-13}$&   $1.07 \times 10^{-11}$  &$4.51\times 10^{-10}$  &1.8&1.8&1.1&200.0\\
%Run 14& $6\times  10^{-13}$&   $1.07 \times 10^{-11}$  &$4.51\times 10^{-10}$  &1.8&1.8&1.1&150.0\\
%Run 15& $6\times  10^{-13}$&   $6    \times 10^{-8}$   &$1.92\times 10^{-1}$              &1.2&1.2&1.1&200.0\\
%Run 16& $6\times  10^{-13}$&   $6    \times 10^{-8}$   &$1.92\times 10^{-1}$             &1.2&1.2&1.1&150.0\\
\hline
\end{tabular}
\label{tab:params}
\end{table*}

\clearpage

\end{document}